\documentclass[12pt]{article}

\newenvironment{numberedlist}
{\begin{list}{\makebox[20pt]{\hss(\arabic{itemno})\enspace}}
             {\usecounter{itemno}\labelwidth 20pt}}{\end{list}}

\newcounter{itemno}

\newcounter{itemno1}

\newcounter{itemno2}

\newcounter{exno}

\newcounter{defno}

\newenvironment{defn}{\refstepcounter{defno}\medskip \noindent {\bf
Definition \thedefno.\ }}{\medskip}

\newcommand{\sep}{\;\vert\;}

\newcommand{\oprove}{\vdash\kern-.6em\lower.7ex\hbox{$\scriptstyle O$}\,}

\newcommand{\Dscr}{{\cal D}}
\newcommand{\Pscr}{{\cal P}}

\newcommand{\pderivation}{{\cal P}\kern -.1em\hbox{\rm -derivation}}
\newcommand{\pderivationl}{{\cal P}\kern -.1em\hbox{\em -derivation}}
\newcommand{\pderivable}{{\cal P}\kern -.1em\hbox{\rm -derivable}}
\newcommand{\pderivablel}{{\cal P}\kern -.1em\hbox{\em -derivable}}
\newcommand{\pderivations}{{\cal P}\kern -.1em\hbox{\rm -derivations}}
\newcommand{\pderivability}{{\cal P}\kern -.1em\hbox{\rm -derivability}}

\newcommand{\all}{\forall}
\newcommand{\some}{\exists}

\newsavebox{\lpartfig}
\newsavebox{\rpartfig}

\newenvironment{exmple}{
 \begingroup \begin{tabbing} \hspace{2em}\= \hspace{3em}\= \hspace{3em}\=
\hspace{3em}\= \hspace{3em}\= \hspace{3em}\= \kill}{
 \end{tabbing}\endgroup}

\newcommand{\lb}{\langle}
\newcommand{\rb}{\rangle}
\newcommand{\pr}{prov}

\newcommand{\prove}{exec} % choice conjunction
     {\\* \hspace*{\fill} \end{trivlist}}

\newcommand{\muprolog}{{Prolog$^{0/1}$}}
\renewcommand{\pr}{pv}
\renewcommand{\prove}{ex} % choice conjunction
\newcommand{\bli}{\all^{b}} %
\newcommand{\defeq}{\mathrel{\stackrel{{\scriptstyle\triangle}}{=}}}
\renewcommand{\Pscr}{{\cal G}}

\begin{document}

\begin{center}
{\Large {\bf Combining Fixed-Point Definitions and Game Semantics in Logic Programming}}
\\[20pt] 
{\bf Keehang Kwon}\\
Dept. of Computer  Engineering, DongA University \\
Busan 604-714, Korea\\
%051-200-7784 \\
  khkwon@dau.ac.kr\\
\end{center}

\noindent {\bf Abstract}: 
Logic programming with fixed-point definitions is a useful extension of
traditional logic programming.  Fixed-point definitions can capture
simple model checking problems and closed-world assumptions.
Its operational semantics is typically based on intuitionistic provability. 

   We extend the  operational
semantics of these languages with game semantics. This extended 
 semantics has several interesting
aspects: in particular, it gives a logical status to
the $read$ predicate.
%\end{summary}

{\bf keywords:} interaction, game semantics, read, computability logic.
%\end{keywords}

\section{Introduction}\label{sec:intro}

Logic programming with fixed-point definitions is a useful extension to the logic
of Horn clauses. In this approach (see, for example, \cite{mcdowell97lics,tiu05eshol}), 
clauses  of the form  $A\defeq B$ -- called $definition$ clauses -- are used to provide
least fixed-point definitions of atoms. We assume that a set $\Dscr$ of such definition 
clauses -- which we call program -- has been fixed.   The following {\em
  definition-right} rule, which is a variant of the one used in 
LINC\cite{tiu05eshol}, is used in this paper
as an inference  rule which introduces atomic
formulas on the right. \\

%\[
%\vcenter{
%  \infer
%      {\Sigma:\Pscr\vdash A}
%      {\Sigma:\\Pscr\vdash B\theta}}
%  \hbox{\ , provided $A'\defeq B\in\Dscr$ and $A'\theta=A$.}
%\]

$\pr(\sigma,\Pscr\vdash A)$ if $A'\defeq B\in\Dscr$ and  $A'\theta=A\sigma$ and
$\pr(\sigma\theta,\Pscr\vdash B)$. \\

This rule is similar to backchaining in Prolog with the difference that an answer
 subsititution $\sigma$ is maintained and 
applied to formulas as $lazily$ as possible here.
 The {\em definition-left} rule is a case analysis in reasoning.\\

%\[
%\infer{\Sigma:\Gamma, A \vdash G}
%      {\{\Sigma\theta:\Gamma\theta,B\theta\vdash G\theta
%       \ | \ A'\defeq B\in\Dscr\mbox{ and }
%       \theta \in csu(A, A')\}}
%\]

$\pr(\sigma, A:\Pscr\vdash D)$ if, for each $\theta$ which is the $mgu(A\sigma, A'
)$ for some $A'\defeq B\in\Dscr$,
    $\pr(\sigma\theta, B:\Pscr\vdash D)$. \\

This rule is well-known and used to instantiate the  free variables of
the sequent by
$\theta$, which is a most general unifier (mgu) for
atoms $A\sigma$ and $A'$.   If there is no such $\theta$, the sequent is proved.

The operational semantics of these languages is typically based on intuitionistic provability.
In the operational semantics based on  provability,
solving the universally quantified goal $\all x D$ from a definition $\Dscr$
 simply {\it terminates} with a success if it is provable.

In this paper, we  make the above  operational semantics more ``constructive'' and ``interactive'' 
by adopting the game semantics in \cite{Jap03,Jap08}.
That is, our approach in this paper involves a modification of the  operational
semantics to allow for more active participation from the user.
Solving  $\all x D$ from a program $\Dscr$ now  has the
following two-step operational semantics:

\begin{itemize}

\item Step (1): the machine tries to prove $\all x D$ from a program $\Dscr$. If it fails, the
machine returns the failure. If it succeeds, goto Step (2).

\item Step (2): the machine requests the user to choose a constant $c$ for $x$ and then proceeds
with solving the  goal, $[c/x] D$.

\end{itemize}

As an   
illustration of this approach, let us consider the following program. \\

$ \{\ emp(tom) \defeq \top.\ emp(pete)\defeq\top.\ $

$   boss(tom,bob) \defeq \top.\ boss(pete,bob)\defeq\top.\ $

$ wife(tom,mary)\defeq\top.\ wife(pete,ann)\defeq\top.\ wife(john,sue)\defeq\top.\ \} $ \\

\noindent 
 As a particular example, consider a goal task 
 $\all x ( emp(x) \supset  \exists y\   wife(x,y))$.
 This goal simply terminates
with a success in the context of \cite{tiu05eshol} as it is solvable.  However, in our 
context,
  execution requires more. To be specific, execution proceeds as follows: the system 
 requests the user to select a particular employee for $x$. 
After the   employee  -- 
 say, $tom$ -- is selected, the system returns $y = mary$.
     As seen from the example above, universally quantified goals in intuitionistic logic
can be used to model the $read$ predicate in Prolog.

We also introduce $blind$ universal quantifiers of the form $\bli x D$. 
This quantification  is similar to $\all x D$ but is read as ``for an unknown value for $x$''.
The machine therefore does $not$ request the user to choose any value for $x$ for this
quantification.
As an   
illustration of this quantifier, let us consider  a goal task $\some y \bli
 x (emp(x) \supset boss(x,y))$.
In this case, execution proceeds as follows: the system chooses $bob$ for $y$ and then
successfully terminates without requesting the user to choose a value for $x$. 
  
In this paper we present the syntax and semantics of this  language called \muprolog. 
The remainder of this paper is structured as follows. We describe a subset of LINC
  logic  in
the next section.  Section 3 describes the new semantics.
Section 4 concludes the paper. 

\section{An Overview of  Level 0/1 prover}

Our language is  a variant  of a subset of the level 0/1 prover in \cite{tiu05eshol},  which is a 
simple  fragment of LINC. Therefore, we closely follow their presentation in 
\cite{tiu05eshol}.
The  language can also be seen as a version of Horn clauses
 with some extensions. We assume that  a program -- a set of definition clauses $\Dscr$ --
is given. 
We have two kinds of goals given  
by $G$- and $D$-formulas  below:
\begin{exmple}
\>$G ::=$ \>  $\top \sep \bot \sep A \sep   G \land  G \sep  G \lor  G \sep  \some x\ G   $ \\   \\
\>$D ::=$ \>  $\top \sep \bot \sep A  \sep D \land D \sep   D \lor D \sep  \some x\ D  \sep \all x\ D \sep \bli x\ D \sep G \supset D $\\
\end{exmple}
\noindent
In the rules above, $A$  represents an atomic formula.

The formulas in this languages are divided into {\em level-0} goals, 
given by $G$ above, and {\em level-1} goals, given by $D$. 
We assume that atoms are partitioned
 level-0 atoms and level-1 atoms. Goal formulas can be level-0 
or level-1 formulas, and in a definition $A \defeq B$, $A$ and $B$ can
be level-0 or level-1 formulas, provided that  level($A$)
$\geq$   level($B$). 

Level-0 formulas and Level-1 formulas are similar to
goal formulas in Prolog.
However, when the Level-1 prover meets the implication $G\supset D$, it
attempts to solve $G$. If $G$ is solvable with  all the possible  answer
substitutions $\sigma_1,\ldots,\sigma_n$, then the Level-1 prover  checks that, for every 
substitution $\sigma_i$, $D\sigma$ holds.
If Level-0 finitely fails, the implication is proved.

 We will  present the standard operational 
semantics for this language  as inference rules \cite{Khan87}. 
 Below the notation $G:\Pscr$ denotes
$\{ G \} \cup \Pscr$. Note that execution  alternates between 
two phases: the left rules phase
and the right rules phase.
In this fragment, all the left rules are 
invertible and therefore  the left-rules take precedence over the right rules.
Note that our semantics is a lazy version of the semantics of level 0/1 prover in the sense that
an answer substitution is applied as lazily as possible.  This makes it easy to
 transit smoothly to the game-based execution model in the next section.

\begin{defn}\label{def:semantics}
Let $\sigma$ be an answer substitution and let $G,D$ be a goal and let $\Pscr$ be a set of
$G$-formulas.
Then the task of proving $D$ from an empty set with respect to $\sigma,\Dscr$ -- 
$\pr(\sigma, \emptyset\vdash D)$ (level 1)-- and the task of proving $D$ from $\Pscr$ with respect to $\sigma,\Dscr$ -- $\pr(\sigma, \Pscr \vdash D)$ (level 0)--
are (mutual recursively) defined as 
follows:

\begin{numberedlist}

% Below is the description of the   level-0 prover

\item  $\pr(\sigma, \bot:\Pscr\vdash D)$. \% This is a success.

\item $\pr(\sigma, \top:\Pscr\vdash D)$ if
    $\pr(\sigma, \Pscr \vdash D)$. \% $\top$ in the premise is redundant.

\item $\pr(\sigma, A:\Pscr\vdash D)$ if, for each $\theta$ which is the $mgu(A\sigma, A')$ for some $A'\defeq B\in\Dscr$,
    $\pr(\sigma\theta, B:\Pscr\vdash D)$. \% DefL rule

\item    $\pr(\sigma, G_0\land G_1:\Pscr\vdash D)$ if   $\pr(\sigma, G_0:G_1:\Pscr\vdash D)$. 

\item    $\pr(\sigma, G_0\lor G_1:\Pscr\vdash D)$ if   $\pr(\sigma, G_0:\Pscr\vdash D)$ 
and $\pr(\sigma, G_1:\Pscr\vdash D)$. 

\item    $\pr(\sigma, \some x G:\Pscr\vdash D)$ if   $\pr(\sigma, [y/x]G:\Pscr\vdash D)$ where $y$ is a $new$ free variable. 

\% Below is the description of the   level-1 prover

\item    $\pr(\sigma, \emptyset\vdash \top)$. \% solving a true goal

\item $\pr(\sigma, \emptyset\vdash A)$ if $A'\defeq B\in\Dscr$ and  $A'\theta = A\sigma$ and
$\pr(\sigma\theta, \emptyset\vdash B)$. \% DefR

\item $\pr(\sigma, \emptyset\vdash D_0 \land D_1)$  if $\pr(\sigma, \emptyset\vdash D_0)$ and 
$\pr(\sigma, \emptyset\vdash D_1)$. 

\item $\pr(\sigma, \emptyset\vdash D_0 \lor D_1)$  if $\pr(\sigma, \emptyset\vdash D_i)$ where $i$ is 0 or 1.

 \item $\pr(\sigma, \emptyset\vdash G \supset D)$  if $\pr(\sigma, G:\emptyset\vdash D)$. \% switch from level 1 to level 0

 \item $\pr(\sigma, \emptyset\vdash \all x D)$  if $\pr(\sigma, \emptyset\vdash [y/x]D)$ where $y$ is a $new$ free variable.

\item $\pr(\sigma, \emptyset\vdash \bli x D)$  if $\pr(\sigma, \emptyset\vdash [y/x]D)$ where $y$ is a $new$ free variable.

\item $\pr(\sigma, \emptyset\vdash \some x D)$  if $\pr(\sigma\sigma_1, \emptyset\vdash\ [w/x]D)$ 
where $w$ is a new free variable, $\sigma_1 =  \{\lb w,t\rb \}$ and $t$ is a term.

% \item $\pr(\sigma,\Pscr, \some x D)$  if $\pr(\sigma,\Pscr, [y/x]G)$ 
%where $y$ is a new free variable.

\end{numberedlist}
\end{defn}
\noindent Most rules are straightforward to read.

The following is a proof tree of the example given in Section 1. 
Below the proof tree is represented  as a list.
Now, given  $\sigma,\Pscr$ and  $D$, a proof tree of a {\it proof formula}
$(\sigma,\Pscr,D)$  is a list of tuples of
the form $\lb E,Ch \rb$ where $E$ is a proof formula and $Ch$ is  a list
of the form $i_1::\ldots::i_n::nil$ where each $i_k$
is the address of
its $k$th child (actually the distance to $E$'s $k$th chilren
in the proof tree). \\

$\{ (h_0,tom),(w_0,ann) \}$, $\emptyset\vdash \top$, nil \% success

$\{ (h_0,tom),(w_0,ann) \}$, $\emptyset\vdash wife(h_0,w_0)$, 1::nil \% defR

$\{ (h_0,pete) \}$, $\emptyset\vdash \some y\ wife(h_0,y)$, 1::nil \% $\some$-R

$\{ (h_0,tom),(w_0,mary) \}$, $\emptyset \vdash \top$, nil \% success

$\{ (h_0,tom),(w_0,mary) \}$, $\emptyset \vdash wife(h_0,w_0)$, 1::nil \% defR

$\{ (h_0,tom) \}$, $\emptyset \vdash \some y\ wife(h_0,y)$, 1::nil \% $\some$-R

$\emptyset$, $emp(h_0) \vdash \some y\ wife(h_0,y)$, 4::1::nil \% defL

$\emptyset$, $\emptyset \vdash emp(h_0) \supset \some y\ wife(h_0,y)$, 1::nil

$\emptyset$, $\emptyset \vdash \all x  ( emp(x) \supset \some y\ wife(x,y))$, 1::nil \% $\all$-R \\

\section{An Alternative Operational Semantics}\label{sec:0627}

Adding game semantics  requires two execution phases: (1) the proof phase and (2) 
the execution phase.
To be precise, our new execution model -- adapted from \cite{Jap03} -- 
 actually solves the goal relative to the program using the proof tree built in the proof
phase.

 In the execution phase, to deal with the universally quantified goals properly,
the machine needs to maintain an $input$ $substitution$ $F$ of the form
$\{ y_0/c_0,\ldots,y_n/c_n \}$ where each $y_i$ is a variable introduced by a universally quantified goal
in the proof phase and each $c_i$ is a constant typed by
the user during the execution phase.
%We also need to maintain a set $H$ of free variables introduced by blind universal quantifiers
%to ensure that these variables which might have been instantiated during the proof phase
%mustmust be uninstantiated in the execution phase. We call $H$ the $blind$ set.

\begin{defn}\label{def:exec}
Let $i$ be an index,  let $L$ be a proof tree, let  $F$ be an input substitution.
% and let $H$ be a blind set.  
Then  executing $L_i$ (the $i$ element in $L$) with $F$ -- written as $\prove(i,L,F)$ --
 is defined as follows: 

\begin{numberedlist}

\item  $\prove(i,L,F)$ if $L_i = (E,nil)$. \% no child, success.

\item  $\prove(i,L,F)$ if $L_i = (\sigma,\emptyset\vdash D_0 \land D_1, m::1::nil)$ and
 $\prove(i-m,L,F)$ and $\prove(i-1,L,F)$. \% two children

\item  $\prove(i,L,F)$ if $L_i = (\sigma,G_0\lor G_1:\Pscr\vdash D, m::1::nil)$ and
 $\prove(i-m,L,F)$ and $\prove(i-1,L,F)$. \% two children

\item  $\prove(i,L,F)$ if $L_i = (\sigma,\emptyset\vdash \all x D, 1::nil)$ and 
$L_{i-1} = (\sigma,\emptyset\vdash [y/x] D, Ch)$
  and $read(k)$ and $\prove(i-1,L, F \cup \{ y/c \})$
 where $c$ is the user input (the value 
stored in $k$). \% update $F$ for universal quantifiers.

%\item  $\prove(i,L,F)$ if $L_i = (\sigma,\emptyset\vdash \bli x D, 1::nil)$ and 
%$L_{i-1} = (\sigma,\emptyset\vdash [y/x] D, n)$
%  and  $\prove(i-1,L, F )$
 
\item  $\prove(i,L,F)$ if $L_i = (\sigma, \emptyset \vdash \some x D, 1::nil)$ and 
$L_{i-1} = (\sigma, \emptyset \vdash [w/x] D, Ch)$
  and $print(x = w \sigma F)$ and $\prove(i-1,L, F \cup \{ y/c \})$

\item  $\prove(i,L,F)$ if $L_i = (\sigma,A:\Pscr\vdash D, i_1::\ldots::i_n::nil)$ and 
choose a $i_k$ such that $L_{i-i_k} = (\sigma\theta_k,B:\Pscr\vdash D, Ch)$ and ($F$ and $\theta_k$ agree on the variables
appearing in $F$)
  and $\prove(i-i_k,L, F)$. \% choose a correct one among many paths in defL

%\item  $\prove(i,L,F,H)$ if $L_i = (\sigma,\Pscr, \some x G, 1)$ and 
%$L_{i-1} = (\sigma\{ \lb y,t \rb \}\sigma_1,\Pscr, [y/x] G, n)$
%  and $\prove(i-1,L',F,H)$ where  $x = yF\sigma_H$ . 
% Hence the value of $x$ is $y$ instantiated by $F$ and $\sigma_H$.

\item  $\prove(i,L,F)$ if $L_i = (E,1::nil)$  and $\prove(i-1,L,F)$. \% otherwise

\end{numberedlist}
\end{defn}

\noindent  
Initially, $F$  is an empty substitution.
The following is an execution sequence of the goal
$\all x  ( emp(x) \supset \some y\ wife(x,y))$ using the  proof tree above. We assume  here
that the user chooses $pete$ for $x$.
Note that the last  component represents $F$. \\

$\{ (w_0,ann) \}$, $\top\ \vdash \top$, $\{ (h_0,pete) \}$ \% success

$\{ (w_0,ann) \}$, $\top\ \vdash wife(h_0,w_0)$, $\{ (h_0,pete) \}$ \% defR

$\emptyset$, $\top\ \vdash\ \some y\ wife(h_0,y)$, $\{ (h_0,pete) \}$  \% $\some$-R

$\emptyset$, $emp(h_0)\ \vdash\ \some y\ wife(h_0,y)$, $\{ (h_0,pete) \}$   \% defL

$\emptyset$, $\emptyset\ \vdash\ emp(h_0) \supset \some y\ wife(h_0,y)$, $\{ (h_0,pete) \}$ \% update $F$

$\emptyset$, $\emptyset\ \vdash\ \all x  ( emp(x) \supset \some y\ wife(x,y))$, $\emptyset$ \% $\all$-R \\

\section{Conclusion}\label{sec:conc}

In this paper, we have considered a new execution model for the level 0/1 prover.
 This new model is interesting in that it gives a logical status to the $read$ predicate in Prolog.
We plan to connect our execution model to Japaridze's Computability Logic \cite{Jap03,Jap08}
 in the near future.

%\section{Acknowledgements}

%This work  was supported by Dong-A University Research Fund.

\bibliographystyle{plain}

%\profile*{}{}% without picture of author's face

%\end{multicols}

\end{document}